   \documentclass[]{article}
  \usepackage{amsmath, amsthm, amssymb,graphics,epsf,authblk,wrapfig}
  \usepackage[round]{natbib}
  \usepackage{parskip}

\newcommand{\code}[1]{\texttt{#1}}

  \usepackage{amsmath, amsthm, amssymb,graphics,epsf,epstopdf,graphicx,geometry,mathrsfs}
  \usepackage{listings}
  \usepackage{listings}
\usepackage{color}
\usepackage{cancel}
\usepackage{hyperref}
 
\definecolor{dkgreen}{rgb}{0,0.6,0}
\definecolor{gray}{rgb}{0.5,0.5,0.5}
\definecolor{mauve}{rgb}{0.58,0,0.82}
 
\title{Pyrcca: regularized kernel canonical correlation analysis in Python and its applications to neuroimaging.}  
\author[1]{Natalia Y. Bilenko}
\author[1,2]{Jack L. Gallant}
\affil[1]{Helen Wills Neuroscience Institute, University of California, Berkeley}
\affil[2]{Department of Psychology, University of California, Berkeley}

\begin{document}
\maketitle

\begin{abstract}

Canonical correlation analysis (CCA) is a valuable method for interpreting cross-covariance across related datasets of different dimensionality. There are many potential applications of CCA to neuroimaging data analysis. For instance, CCA can be used for finding functional similarities across fMRI datasets collected from multiple subjects without resampling individual datasets to a template anatomy. In this paper, we introduce Pyrcca, an open-source Python module for executing CCA between two or more datasets. Pyrcca can be used to implement CCA with or without regularization, and with or without linear or a Gaussian kernelization of the datasets. We demonstrate an application of CCA implemented with Pyrcca to neuroimaging data analysis. We use CCA to find a data-driven set of functional response patterns that are similar across individual subjects in a natural movie experiment. We then demonstrate how this set of response patterns discovered by CCA can be used to accurately predict subject responses to novel natural movie stimuli.
\end{abstract}

\section{Introduction}
Covariance analyses are regarded as one of the simplest approaches for finding similarities across datasets. One type of covariance analysis, known as canonical correlation analysis (CCA), is a commonly used method in statistics. It was first introduced by \citet{Hotelling1936} as a method for finding relationships between two sets of variables, and in the decades that followed, it has been applied in a variety of scientific fields, including neuroscience \citep{Hardoon2007}. For instance, in neuroimaging, CCA in conjunction with independent component analysis (ICA) has been used to find networks of brain activity in resting state experiments \citep{Varoquaux2010}. 

Covariance and correlation methods have been applied previously to find cross-subject mapping in neuroimaging experiments. One such method, functional timecourse correlation, has been used to find brain activity networks or similarities across individual subjects, particularly in a technique called \textit{hyperalignment} \citep{Haxby2011}. Hyperalignment aligns fMRI datasets in the functional space, instead of the anatomical space, and has been shown to be more effective for between-subject classification using multi-voxel pattern analysis (MVPA) than anatomical alignment. However, this method is constrained to comparing parameter spaces of equal dimensionality. Additionally, correlation methods quantify similarity between datasets, but make it very challenging to interpret how exactly these datasets might be related to each other. In contrast to timecourse correlation, CCA is not limited to datasets of corresponding dimensionality. It also reveals interpretable components of input data in addition to quantifying their similarity.

There are several software packages available for implementing CCA in MATLAB, including a CCA-fMRI toolbox in SPM, a well-known fMRI analysis software package \citep{cca-fmri}. While a few packages in Python for implementation of CCA have been developed \citep{scikit-learn, pykcca}, they have either been very minimal or do not seem to be undergoing active development. Here we introduce a Python package for implementation of CCA called Pyrcca, which stands for PYthon Regularized Canonical Correlation Analysis. We provide options for kernel and non-kernel CCA and regularization, and for cross-validation of two hyperparameters: regularization coefficient and number of canonical variates. We then demonstrate the use of Pyrcca for an application of CCA in fMRI analysis, by finding similar dimensions of brain activity across subjects in a natural movie experiment.

\subsection{Canonical Correlation Analysis}

Canonical correlation analysis (CCA) is a statistical method for finding correlational linear relationships between two or more multidimensional variables \citep{Hotelling1936, Hardoon2004}. CCA finds a pair of bases for each set of the variables, such that linear transformations of each variable onto these bases are maximally correlated. From an information theoretic point of view, CCA maximizes the mutual information between the linear transformations of the datasets.

Given a zero-mean $d \times n$-dimensional data matrix $\mathbf{X} = (x_1, x_2, \dots x_n) \in \mathbb{R}^{d \times n}$ and zero-mean $d \times m$-dimensional data matrix $\mathbf{Y} = (y_1, y_2, \dots y_m) \in \mathbb{R}^{d \times m}$, CCA finds a pair of vectors $\mathbf{u}$ and $\mathbf{v}$, called \textit{canonical variates}, that are linear transformations of $\mathbf{X}$ and $\mathbf{Y}$: $\mathbf{u} = \langle \mathbf{a}, \mathbf{X} \rangle$ and $\mathbf{v} = \langle \mathbf{b}, \mathbf{Y} \rangle$, such that the correlation between these transformations is maximized:
\begin{equation}
\rho = \max \frac{\langle \mathbf{u}, \mathbf{v} \rangle}{\|\mathbf{u}\| \|\mathbf{v}\|}
\end{equation}

Once the first pair of canonical variates is determined, subsequent canonical variates can be found by maximizing the correlation analogously, with the constraint that they are uncorrelated with the preceding canonical variates. The total number of canonical variates must be less or equal to $\min \{n, m\}$.

The expression for $\rho$ can be rewritten in terms of the sample covariance of $\mathbf{X}$ and $\mathbf{Y}$:

\begin{equation}
\rho = \max \frac{\mathbf{a}' C_{\mathbf{XY}} \mathbf{b}}
{\sqrt{\mathbf{a}' C_{\mathbf{XX}} \mathbf{a} \mathbf{b}' C_{\mathbf{YY}} \mathbf{b}}}
\end{equation}

This problem is equivalent to maximizing the numerator, subject to constraints $\mathbf{a}' C_{\mathbf{XX}} \mathbf{a} = 1$ and \\$\mathbf{b}' C_{\mathbf{YY}} \mathbf{b} = 1$.

The resulting Lagrangian is:
\begin{equation}
L(\lambda, \mathbf{a}, \mathbf{b}) = \mathbf{a}'C_{\mathbf{XY}}\mathbf{b} - \frac{\lambda_X}{2}(\mathbf{a}'C_{\mathbf{XX}}\mathbf{a} - 1) - \frac{\lambda_Y}{2}(\mathbf{b}'C_{\mathbf{YY}}\mathbf{b} - 1)
\end{equation}

which reduces to solving the following generalized eigenvalue problem:

\begin{equation}
\left ( \begin{array}{cc} 0 & C_{\mathbf{XY}} \\ C_{\mathbf{YX}} & 0 \end{array} \right) \left ( \begin{array}{c} \mathbf{a} \\ \mathbf{b} \end{array} \right) = \rho^2 \left ( \begin{array}{cc} C_{\mathbf{XX}} & 0 \\ 0 & C_{\mathbf{YY}} \end{array} \right)
\end{equation}

where $\rho$ is the vector of canonical correlations.

\subsubsection{Extension to multiple datasets.}

CCA is easily extended to multiple datasets. The generalized eigenvalue problem becomes:

\begin{equation}
\left ( \begin{array}{ccc} 0 & C_{\mathbf{XY}} & C_{\mathbf{XZ}} \\ C_{\mathbf{YX}} & 0 & C_{\mathbf{YZ}} \\ C_{\mathbf{ZX}} & C_{\mathbf{ZY}} & 0 \end{array} \right) \left ( \begin{array}{c} \mathbf{a} \\ \mathbf{b} \\ \mathbf{c} \end{array} \right) = \rho^2 \left ( \begin{array}{ccc} C_{\mathbf{XX}} & 0 & 0 \\ 0 & C_{\mathbf{YY}} & 0 \\ 0 & 0 & C_{\mathbf{ZZ}} \end{array} \right)
\end{equation}

\subsubsection{Kernelization and regularization of CCA.}

Kernel CCA can be used for additional flexibility, in order to map the data into a high-dimensional space before performing CCA (known as the kernel trick). If the kernel function is nonlinear, kernel CCA can be used to capture nonlinear relationships in the data. To perform kernel CCA, first a kernel function $\phi(\mathbf{X})$ is chosen and the data are projected onto it:

$\phi: \mathbf{X} = (x_1, x_2, \dots x_n) \to \phi(\mathbf{X}) = (\phi_1(\mathbf{X}), \phi_2(\mathbf{X}), \dots, \phi_K(\mathbf{X}))$, where $n < K$.

CCA is then applied in the same manner on the kernel projections of the data, $K_{\mathbf{X}}$ and $K_{\mathbf{Y}}$ by finding linear transformations of $K_{\mathbf{X}}$ and $K_{\mathbf{Y}}$:

\begin{align*}
\mathbf{u} & = \langle \mathbf{a}, K_{\mathbf{X}} \rangle\\
\mathbf{v} & = \langle \mathbf{b}, K_{\mathbf{Y}} \rangle
\end{align*}

However, if the kernels $K_{\mathbf{X}}$ and $K_{\mathbf{Y}}$ are invertible, a perfect correlation can be formed by setting $\mathbf{a}$ to $1$, and solving $\mathbf{b} = (1/\lambda)K_{\mathbf{Y}}^{-1}K_{\mathbf{X}}$. To avoid this trivial solution issue, kernel CCA must be regularized. The regularization is accomplished by penalizing the norms of the weights $\mathbf{a}$ and $\mathbf{b}$, analogously to the regularization procedure used in partial least squares regression. The maximization problem becomes:

\begin{equation}
\rho = \max \frac{\mathbf{a}' K_{\mathbf{X}} K_{\mathbf{Y}} \mathbf{b}}
{\sqrt{(\mathbf{a}' K_{\mathbf{X}}^2 \mathbf{a} + \kappa \|a\|^2) \cdot (\mathbf{b}' K_{\mathbf{Y}}^2 \mathbf{b} + \kappa\|b\|^2)}}
\end{equation}

where $\kappa$ is the regularization parameter.

Regularization can also be used in non-kernel CCA if $d < \min{m, n}$, resulting in an ill-posed problem. If $d > \min{m, n}$, non-kernel CCA can be performed without regularization.

Adjusting the Lagrangian and solving it results in the following generalized eigenvalue problem:

\begin{equation}
\left ( \begin{array}{cc} 0 & K_{\mathbf{X}}K_{\mathbf{Y}} \\ K_{\mathbf{Y}}K_{\mathbf{X}} & 0 \end{array} \right) \left ( \begin{array}{c} \mathbf{a} \\ \mathbf{b} \end{array} \right) = \rho^2 \left ( \begin{array}{cc} K_{\mathbf{X}}^2+\kappa I & 0 \\ 0 & K_{\mathbf{Y}}^2+\kappa I \end{array} \right)
\end{equation}
where $\rho$ is the vector of canonical correlations.

\subsubsection{Comparison with partial least squares regression.}
Mathematically, CCA is very similar to partial least squares regression (PLS). The maximization problem solved in PLS is:
\begin{equation}
\rho = \max \frac{\mathbf{a}' C_{\mathbf{XY}} \mathbf{b}}
{\sqrt{\mathbf{a}' \mathbf{a} \mathbf{b}' \mathbf{b}}}
\end{equation}

This can be reformulated as an eigenvalue problem, similarly to CCA:
\begin{equation}
\left ( \begin{array}{cc} 0 & C_{\mathbf{XY}} \\ C_{\mathbf{YX}} & 0 \end{array} \right) \left ( \begin{array}{c} \mathbf{a} \\ \mathbf{b} \end{array} \right) = \rho^2 \left ( \begin{array}{cc} I & 0 \\ 0 & I \end{array} \right)
\end{equation}

Thus, the problem is virtually identical to CCA, but does not include normalization by autocovariances of the data matrices. It can be thought of as asymptotically large regularization of CCA, where $\kappa I >> C_{XX}, C_{YY}$.

\subsection{Previous implementations of CCA in Python}
\subsubsection{CCA in scikit-learn.}
Scikit-learn \citep{scikit-learn} is a popular machine learning package in Python that includes a variety of algorithms for data mining and analysis. It is developed by many contributors, primarily based at INRIA, France. The implementation of CCA in scikit-learn is in the cross-decomposition module, It inherits from the PLS module, since the two methods are mathematically similar, as described above. This implementation is effective for standard CCA, where an orthogonal set of components needs to be identified for datasets, where the number of samples exceeds the number of features. However, it does not allow for regularization or for the kernel extension of CCA.

\subsubsection{PyKCCA.}
PyKCCA \citep{pykcca} is a Python module for CCA available on GitHub, developed by Lorenzo Riano. It allows for the implementation of kernel CCA. This software does not seem to be under active development, and there are no simple start-up instructions, making it difficult to access for naive users.

\section{Materials \& Methods}
Pyrcca is a module in Python that can be downloaded from GitHub (\url{http://github.com/gallantlab/pyrcca}). It uses functions from the standard Python library, NumPy, SciPy, and Pytables. We kept library dependencies minimal to simplify the module for users with limited programming experience.

\subsection{Pyrcca module organization}
The Pyrcca module is simply organized, with all methods and classes defined in one file: rcca.py. There are two main object classes: \code{rcca.CCA} and \code{rcca.CCACrossValidate}, both of which inherit from a common parent class, \code{rcca.\_CCABase}. The first object class, \code{rcca.CCA}, is intended for implementing the analysis for one set of specific parameters. The second object class, \code{rcca.CCACrossValidate}, is intended for performing cross-validation with a set of parameters and choosing the best solution. There is a basic example of CCA implementation using Pyrcca below, followed by a detailed description of the moduele and a visual representation in Figure \ref{fig:01}.

\subsubsection{A simple example.}
In this example, we create two random datasets with two latent variables, and use Pyrcca to implement CCA between them. The datasets are broken up into two halves. First, we use the first half of the datasets to train a CCA mapping. Then, we test the found mapping we found by validating it on the second half of the datasets. This procedure assures that the found canonical variates are generalizable and are not overfitting to the training data.

You can also explore this example interactively in an IPython notebook included in the GitHub repository: \url{http://github.com/gallantlab/pyrcca}.

\begin{lstlisting}
import numpy as np
import rcca

nObservations = 1000

# Define two latent variables
lat_var1 = np.random.randn(nObservations,)
lat_var2 = np.random.randn(nObservations,)

# Define independent signal components
indep1 = np.random.randn(nObservations, 4)
indep2 = np.random.randn(nObservations, 4)

# Define two datasets as a combination of latent variables
# and independent signal components
data1 = indep1 + np.vstack((lat_var1, lat_var1, lat_var2, lat_var1)).T
data2 = indep2 + np.vstack((lat_var1, lat_var1, lat_var2, lat_var1)).T

# Divide data into two halves: training and testing sets
train1 = data1[:nObservations/2]
test1 = data1[nObservations/2:]
train2 = data2[:nObservations/2]
test2 = data2[nObservations/2:]

# Set up Pyrcca
cca = rcca.CCA(kernelcca=False, numCC=2, reg=0.)

# Find canonical components
cca.train([train1, train2])

# Test on held-out data
corrs = cca.validate([test1, test2])
\end{lstlisting}

\subsubsection{Analysis overview.}
As the first step in the analysis, a CCA object is initialized. CCA can be initialized with or without kernelization, and either a linear or a Gaussian kernel may be used. The other parameters set at initialization are the number of canonical components to be retained and the regularization parameter. Regularization must be applied if the kernel is used, and may be applied if no kernel is used. These two parameters, regularization parameter and the number of canonical components retained, can either be set to specific values, in which case the \code{rcca.CCA} object class should be used, or they can be set to ranges of values to be cross-validated, in which case the \code{rcca.CCACrossValidate} object class should be used to find the best solution.

The \code{train()} method is then employed to find a CCA mapping for two or more supplied datasets, returning sets of canonical components, canonical weights, and canonical correlations. If the goal of the researcher is to quantify dataset similarity or to find the underlying space of correlated components, this completes the analysis. If the researcher's additional goal is to use the CCA mapping to predict held-out data, the \code{validate()} method is used to accomplish that. The researcher supplies held-out datasets, and this method returns predictions for each dataset based on the CCA method and other held-out datasets, as well as feature correlations between the predictions and the actual data. It is also possible to compute variance explained by the canonical components for each feature in the data, using the \code{compute\_ev()} method.

Finally, the \code{save()} and \code{load()} methods are used for saving the results of the CCA analysis to disk and loading them from disk. We use the \code{cPickle} library to save the results.

Below, the details of each object class, class methods, and shared module methods are described.

\subsubsection{\code{rcca.CCA} object class.}
The object class \code{rcca.CCA} is initialized with the following parameters:

\begin{itemize}
\item \code{reg} - regularization parameter, floating point number. Default value is $0.1$.
\item \code{numCC} - number of canonical components that should be retained in the analysis, integer. Default value is $10$.
\item \code{kernelcca} - indicator for whether or not the kernel version of CCA is used, boolean. Default value is \code{True}.
\item \code{ktype} - type of kernel to be used in the analysis if \code{kernelcca} is \code{True}. Unused if \code{kernelcca} is \code{False}. The value can be set to \code{None} (gets recast to \code{"linear"} if \code{kernelcca} is \code{True}, \code{"linear"}, or \code{"gaussian"}. Default value is \code{"linear"}.
\item \code{cutoff} - optional regularization parameter to perform spectral cutoff when computing the canonical weight pseudoinverse during held-out data prediction, floating point number. Default value is $1e-15$, which is equivalent to no regularization.
\item \code{verbose} - indicator for verbose code output, boolean. Default value is \code{True}.
\end{itemize}

The methods of the object class \code{CCA} are inherited from the parent class \code{\_CCABase} and include \code{CCA.train()}, \code{CCA.validate()}, \code{CCA.compute\_ev()}, \code{CCA.save()}, and \code{CCA.load()}.

The method \code{CCA.train()} accepts an argument \code{data}, which should be a \code{list} of two or more datasets between which the CCA analysis will be fit. The dataset dimensionality should be the number of samples by the number of features, where the number of features can vary across datasets.

The method \code{CCA.validate()} accepts an argument \code{vdata}, which should be a \code{list} of two or more held-out datasets from the same population as the datasets used in training, in the same order. The dimensionality for each dataset should also be the number of samples by the number of features, with the features equivalent to those in the training datasets.

The method \code{CCA.compute\_ev()} accepts an argument \code{data}, which should similarly be a \code{list} of two or more datasets in the similar format to above. This method computes variance explained by the CCA mapping in the supplied dataset features.

The methods \code{CCA.save()} and \code{CCA.load()} are corresponding methods which accept an argument \code{fname}, which should be a name of a .pkl file. The method \code{CCA.save()} saves all attributes of the \code{CCA} object to a \code{cPickle} binary file. The method \code{CCA.load()} is used to load attributes from a \code{cPickle} file into a newly initalized \code{CCA} object.

\subsubsection{\code{rcca.CCACrossValidate} object class.}
The object class \code{rcca.CCACrossValidate} is initialized with the following parameters:

\begin{itemize}
\item \code{regs} - array of regularization parameters to test, \code{numpy.array} or \code{list} of floating point numbers. Default value is \code{numpy.array(numpy.logspace(-3, 1, 10))}.
\item \code{numCCs} - array of numbers of canonical components that should be retained in the analysis, \code{numpy.array} or \code{list} of integers. Default value is \code{numpy.arange(5, 10)}.
\item \code{numCV} - number of cross-validation samples to test in the training procedure, integer. Default value is 10.
\item \code{kernelcca} - indicator for whether or not the kernel version of CCA is used, boolean. Default value is \code{True}.
\item \code{ktype} - type of kernel to be used in the analysis if \code{kernelcca} is \code{True}. Unused if \code{kernelcca} is \code{False}. The value can be set to \code{None} (gets recast to \code{"linear"} if \code{kernelcca} is \code{True}, \code{"linear"}, or \code{"gaussian"}. Default value is \code{"linear"}.
\item \code{cutoff} - optional regularization parameter to perform spectral cutoff when computing the canonical weight pseudoinverse during held-out data prediction, floating point number. Default value is $1e-15$, which is equivalent to no regularization.
\item \code{select} - proportion of prediction performance values that should be used to select the optimal hyperparameter, float. Default value is $0.2$, which means that the hyperparameter is selected based on the mean of the top 20\% of correlation values during cross-validation.
\item \code{verbose} - indicator for verbose code output, boolean. Default value is \code{True}.
\end{itemize}

The methods of the object class \code{rcca.CCACrossValidate} that are inherited from the parent class \code{rcca.\_CCABase} include \code{rcca.CCACrossValidate.validate()}, \\\code{rcca.CCACrossValidate.compute\_ev()}, \code{rcca.CCACrossValidate.save()}, and \\\code{rcca.CCACrossValidate.load()} and are the same as described above for the \code{rcca.CCA} object class.

The method \code{rcca.CCACrossvalidate.train()} differs from the method \code{rcca.CCA.train()}. This version of the algorithm performs cross-validation grid-search on the two hyperparameters: \code{regs} and \code{numCCs} in the ranges provided. For each cross-validation fold, the data are split into two sets (80\% and 20\% of the data respectively, using batch cross-validation. Then CCA mapping is fit using one pair of the hyperparameter values on the larger set and used to predict each of the held-out smaller sets of data based on the others. This is repeated for \code{numCV} times for each pair of the hyperparameters. The results are aggregated, and the pair of the hyperparameters with the best prediction performance is saved as \code{rcca.CCACrossValidate.best\_reg} and \code{rcca.CCACrossValidate.best\_numCC}. Then, the whole training datasets are used to fit the CCA mapping with these chosen hyperparameters. The rest of the analysis proceeds analogously to the analysis for the object class \code{rcca.CCA}.

\subsection{Pyrcca applications to neuroimaging}
We demonstrate results of an application of Pyrcca software to neuroimaging data analysis. We use Pyrcca to run CCA on fMRI responses from three subjects watching a natural movie \citep{Nishimoto2011}. We find a space of canonical components that captures similarities across these fMRI responses. Then, we predict each subject's fMRI responses to a held-out set of natural movies based on the CCA mapping and the other subjects' responses to the held-out stimuli.

\subsubsection{BOLD response similarities across subjects.}
The dataset used here consists of BOLD responses for three subjects who watched natural movies in an fMRI scanner, in three different sessions collected over three separate days. Details of the experiments are described in the original publication about this data \citep{Nishimoto2011}. These data are publicly available on the CRCNS.org database \citep{CRCNS}. These data have been preprocessed using standard motion correction in FSL \citep{FSL1, FSL2, FSL3}, realigned to each subject's anatomical scan, and low-frequency scanner noise has been removed from the data. 

A regularized finite impulse response (FIR) regression model was then estimated for each voxel in each subject's brain. The training dataset consists of 7200 timepoints for each subject. The validation dataset consists of 180 timepoints for each subject, repeated 10 times and then averaged to increase signal to noise. The subjects were scanned with a quarter-head occipital slice prescription, resulting in a 64x64x18 scanning volume covering the occipital lobe, or 73728 voxels. Independent localizer experiments were done to identify retinotopic and motion-selective visual areas in each subject (V1, V2, V3, V3A, V3B, V4, lateral occipital cortex (LO), inferior parietal sulcus (IPS), and middle temporal cortex (MT+)). Only voxels that fell into the regions of interest were retained for the CCA analysis for each subject. This voxel selection resulted in 9137 voxels for subject 1, 8013 voxels for subject 2, and 15122 voxels for subject 3.

Pyrcca was used to compute a kernelized CCA mapping with a linear kernel across the training BOLD responses for all three subjects. The optimal regularization parameter and number of components for each analysis were chosen by cross-validation using the \code{rcca.CCACrossValidate.train()} method. The regularization parameter was chosen among ten logarithmically spaced values between $10e-3$ and $10$. The number of components retained was chosen from a linear range beween 3 and 10 components. Generally, the appropriate range for both the regularization parameter and the optimal number of the components can vary depending on the size of the dataset and the structure of the data, so starting with a wide range and narrowing it later on is recommended.

To evaluate the results, each subject's responses on the validation set were predicted using the computed CCA mapping and the other subjects' responses in each analysis. Correlations between the predicted and actual BOLD responses for each voxel were computed and visualized on the surface of the brain of each subject. Voxelwise canonical weights for each subject were also visualized in order to interpret the computed components. The clips of the movie stimuli corresponding to the highest and lowest weights on each component were identified in order to interpret the response features discovered by CCA. The variance explained by each component for each cortical voxel was computed and visualized on the cortical surface of the subjects.

\section{Results}

\subsection{BOLD response similarities across subjects}

The optimal number of components and the optimal regularization coefficient chosen were 5 components and 0.0077, respectively. Cross-subject mapping via CCA in Pyrcca was then evaluated by examining prediction performance on a held-out dataset using the selected number of components and regularization coefficient. The predicted and actual BOLD responses of the subjects on the held-out set were highly correlated throughout the visual cortex. The mean correlation across voxels for each subject was 0.17, 0.14, and 0.12. The maximum correlation for each subject was 0.72, 0.81, and 0.67. The signficance value for the correlation for each voxel was computed using an asymptotic method, and False Discovery Rate correction was applied. Based on this analysis, the smallest significant correlation value for all subjects was 0.07 (p\textless0.05, corrected for multiple comparisons). The correlations for each voxel for one subject are plotted in Figure \ref{fig:02}. Panel A shows the correlations projected onto the surface of the occipital cortex of the subject. The correlations are plotted in a histogram in Panel B. Correlations averaged across known regions of interest, defined in a separate retinotopy experiment, are plotted in Panel C.

We examined the canonical components of the BOLD responses discovered in the CCA analysis by visualizing the canonical weights for these components on the cortical surface of the subjects. For each canonical component, we identified the movie frames that corresponded to the highest and lowest weights for that component in time, accounting for the hemodynamic function by assuming a peak delay of 5 seconds. Then, we calculated variance explained by each of the canonical components in each cortical voxel and visualized it on the cortical surface. The canonical weights for the first 5 components for one of the subjects in the mapping, the movie frames eliciting highest and lowest canonical weights for each component, and variance explained for each cortical voxel by each of the components are plotted in Figure \ref{fig:03}.

\section{Conclusion}

In this paper, we introduce Pyrcca, a Python module for performing regularized kernel canonical correlation analysis. This module is lightweight and simple to use, and can be easily integrated into a more complex Python-based pipeline. We include a simple cross-validation method for hyperparameter selection, which can be easily parallelized. We then describe a use case of Pyrcca as an application to neuroimaging data analysis. Here we use Pyrcca to find shared features of individual subject fMRI responses to a natural movie experiment. We demonstrate how this framework can be used to predict novel subject responses to a held-out stimulus dataset. The features identified by Pyrcca can be examined for further interpretation.

This application of Pyrcca is just one potential use case in neuroimaging data analysis. Pyrcca can also be used to find shared features between different representations of the same stimulus set, or to find a mapping between the same subject's responses to different experiments. Furthermore, there are many applications of CCA outside of neuroimaging or neuroscience, and we hope that Pyrcca can be seamlessly incorporated into any analysis pipeline to implement CCA.

\subsection{Data Sharing}
Pyrcca software presented in this paper is shared publicly on GitHub: \url{http://github.com/gallantlab/pyrcca}. The fMRI data analyzed in this paper is shared on the public repository CRCNS.org \citep{CRCNS}. The examples described can be found in an IPython notebook in the GitHub repository.

\section*{Acknowledgements}
This work was supported by grants from the National Eye Institute (EY019684) and from the Center for Science of Information (CSoI), an NSF Science and Technology Center, under grant agreement CCF-0939370. NYB was additionally supported by the NSF Graduate Research Fellowship Program. We thank Tolga Cukur, Mark Lescroart, Adam Bloniarz, and Alex Huth for helpful discussions about the analysis and software, and Shinji Nishimoto for sharing the data used in these analyses.

\bibliography{./references.bib}

\section*{Figures}

\begin{figure}
\begin{center}
\includegraphics[width=0.9 \textwidth]{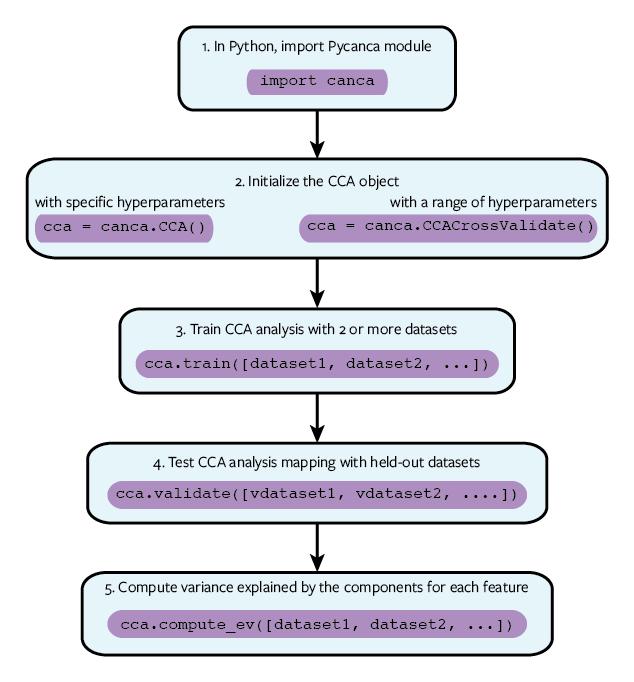}
\end{center}
\textbf{\refstepcounter{figure}\label{fig:01} Figure \arabic{figure}.}{ Overview of the Pyrcca analysis. First, Pyrcca is imported in Python using the command \code{import rcca}. Then, a CCA object is initialized. If specific values of hyperparameters are known, the \code{rcca.CCA} object class is used. If a range of hyperparameter values is being tested, the \code{rcca.CCACrossValidate} object class is used. The CCA mapping is then computed for two or more datasets using the \code{rcca.train()} method. To test the accuracy of the CCA mapping, the method \code{rcca.validate()} is used with two or more held-out datasets. Finally, to compute variance explained by each component for each feature in the data, the method \code{rcca.compute\_ev()} is used. }
\end{figure}

\begin{figure}
\begin{center}
\includegraphics[width=0.9 \textwidth]{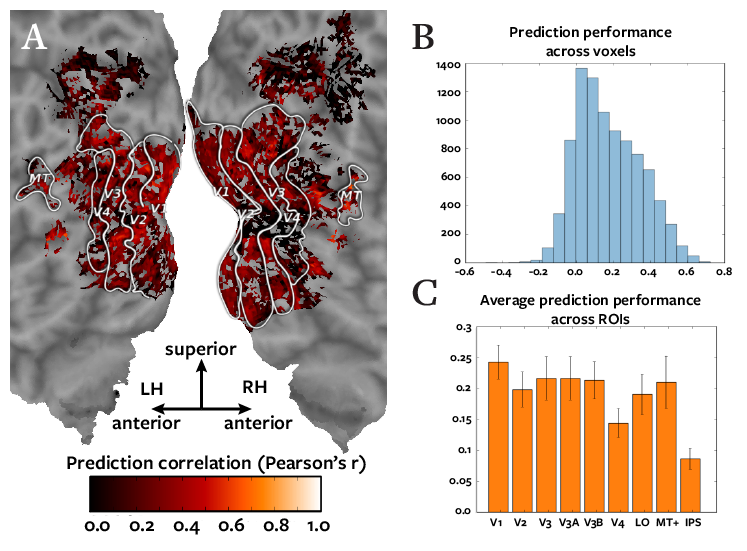}
\end{center}
\textbf{\refstepcounter{figure}\label{fig:02} Figure \arabic{figure}.}{ Prediction performance for a CCA analysis fit between two subject's BOLD responses to a natural movie stimulus. A. Correlations between one subject's actual responses and responses predicted using the other subject's responses and the CCA mapping are plotted on the cortical surface. Each voxel is colored based on the correlation value. Known regions of interest are outlined based on a separate retinotopic mapping experiment. The subject's responses are accurately predicted throughout the visual cortex. B. The described correlations are visualized as a histogram. The mean correlation is 0.17, and the maximum correlation is 0.72. The correlations greater than 0.07 are statistically significant (p\textless0.05, corrected for multiple comparisons). C. This bar plot shows correlations averaged across voxels in specific regions of interest. }
\end{figure}

\begin{figure}
\begin{center}
\includegraphics[width=0.9 \textwidth]{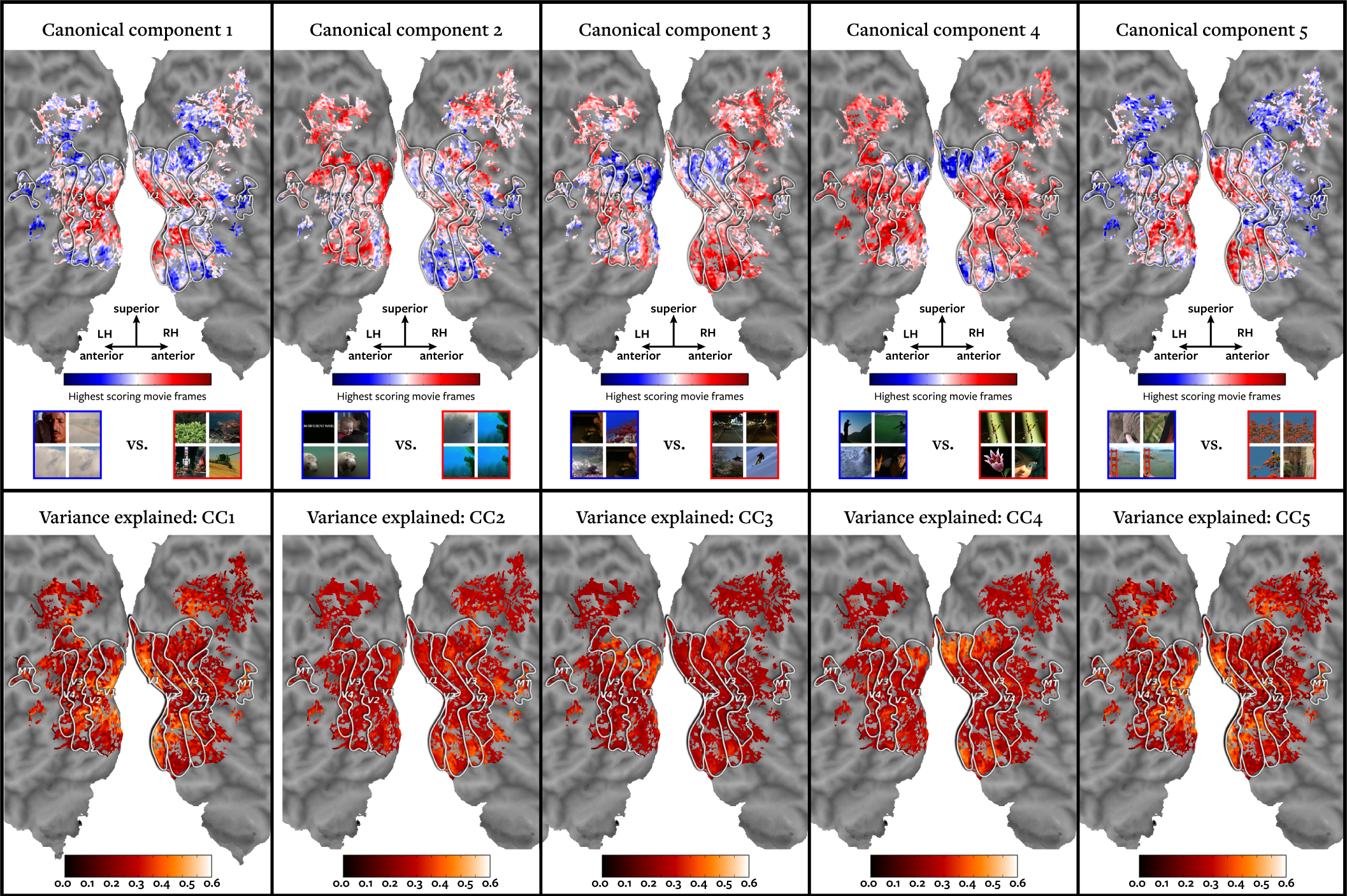}
\end{center}
\textbf{\refstepcounter{figure}\label{fig:03} Figure \arabic{figure}.}{ Five canonical components underlying two subjects' BOLD responses to a natural movie stimulus. The canonical weights for each component are plotted for each voxel on the cortical surface of the subject. Negative weights are indicated in blue, and positive weights are indicated in red. Note that the directionality of each component is arbitrary. The four frames of the movie corresponding to the timepoints with most negative and most positive weights for each component are shown below. Finally, variance explained in each voxel is plotted for each component. Each voxel is colored according to the explained variance value. }
\end{figure}

 \textbf{Figure 1.}{ Overview of the Pyrcca analysis. First, Pyrcca is imported in Python using the command \code{import rcca}. Then, a CCA object is initialized. If specific values of hyperparameters are known, the \code{rcca.CCA} object class is used. If a range of hyperparameter values is being tested, the \code{rcca.CCACrossValidate} object class is used. The CCA mapping is then computed for two or more datasets using the \code{rcca.train()} method. To test the accuracy of the CCA mapping, the method \code{rcca.validate()} is used with two or more held-out datasets. Finally, to compute variance explained by each component for each feature in the data, the method \code{rcca.compute\_ev()} is used. }\label{fig:01}

 \textbf{Figure 2.}{ Prediction performance for a CCA analysis fit between two subject's BOLD responses to a natural movie stimulus. A. Correlations between one subject's actual responses and responses predicted using the other subject's responses and the CCA mapping are plotted on the cortical surface. Each voxel is colored based on the correlation value. Known regions of interest are outlined based on a separate retinotopic mapping experiment. The subject's responses are accurately predicted throughout the visual cortex. B. The described correlations are visualized as a histogram. The mean correlation is 0.17, and the maximum correlation is 0.72. The correlations greater than 0.07 are statistically significant (p\textless0.05, corrected for multiple comparisons). C. This bar plot shows correlations averaged across voxels in specific regions of interest. }\label{fig:02}

 \textbf{Figure 3.}{ Five canonical components underlying two subjects' BOLD responses to a natural movie stimulus. The canonical weights for each component are plotted for each voxel on the cortical surface of the subject. Negative weights are indicated in blue, and positive weights are indicated in red. Note that the directionality of each component is arbitrary. The four frames of the movie corresponding to the timepoints with most negative and most positive weights for each component are shown below. Finally, variance explained in each voxel is plotted for each component. Each voxel is colored according to the explained variance value. }\label{fig:03}

\end{document}